\documentclass[aoas,nameyear,dvips]{arximspdf}
\usepackage{graphics}

\doi{10.1214/09-AOAS282}
\volume{4}
\issue{1}
\pubyear{2010}
\firstpage{366}
\lastpage{395}

\begin{document}
\begin{frontmatter}

\title{Hidden Markov models for alcoholism treatment~trial data}
\runtitle{HMMs for alcohol data}

\begin{aug}
\author[a]{\fnms{Kenneth E.} \snm{Shirley}\corref{}\ead[label=e1]{shirley@stat.columbia.edu}},
\author[b]{\fnms{Dylan S.} \snm{Small}\ead[label=e2]{dsmall@wharton.upenn.edu}},
\author[c]{\fnms{Kevin G.} \snm{Lynch}\ead[label=e3]{lynch\_k@mail.trc.upenn.edu}},\\
\author[d]{\fnms{Stephen A.} \snm{Maisto}\ead[label=e4]{samaisto@syr.edu}}
\and
\author[e]{\fnms{David W.} \snm{Oslin}\ead[label=e5]{oslin@mail.med.upenn.edu}}
\runauthor{K. Shirley et al.}
\affiliation{Columbia University, University of Pennsylvania,
University of Pennsylvania, Syracuse University and University of Pennsylvania}
\address[a]{K. E. Shirley\\Department of Statistics\\
Columbia University\\
1255 Amsterdam Avenue\\
New York, New York 10025\\USA\\
\printead{e1}}

\address[b]{D. S. Small\\
Department of Statistics\\
Wharton School\\ University of Pennsylvania\\
400 Jon M. Huntsman Hall\\
3730 Walnut Street\\ Philadelphia, Pennsylvania 19104-6340\\USA\\ \printead{e2}}

\address[c]{K. G. Lynch\\
Treatment Research Center\\ University of Pennsylvania\\
3900 Chestnut Street\\ Philadelphia, Pennsylvania 19104\\USA\\ \printead{e3}}

\address[d]{S. A. Maisto\\
Department of Psychology\\
414 Huntington Hall\\ Syracuse University\\ Syracuse, New York
13244\\USA\\ \printead{e4}}

\address[e]{D. W. Oslin\\
Department of Psychiatry\\
Department of Veterans Affairs\\
University of Pennsylvania\\
3535 Market Street, Room 3002\\
Philadelphia, Pennsylvania 19104\\USA\\
\printead{e5}}
\end{aug}

\received{\smonth{5} \syear{2007}}
\revised{\smonth{7} \syear{2009}}

%
\begin{abstract}
In a clinical trial of a treatment for alcoholism, a common response
variable of interest is the number of alcoholic drinks consumed by each
subject each day, or an ordinal version of this response, with levels
corresponding to abstinence, light drinking and heavy drinking. In
these trials, within-subject drinking patterns are often characterized
by alternating periods of heavy drinking and abstinence. For this
reason, many statistical models for time series that assume steady
behavior over time and white noise errors do not fit alcohol data well.
In this paper we propose to describe subjects' drinking behavior using
Markov models and hidden Markov models (HMMs), which are better suited
to describe processes that make sudden, rather than gradual, changes
over time. We incorporate random effects into these models using a
hierarchical Bayes structure to account for correlated responses within
subjects over time, and we estimate the effects of covariates,
including a randomized treatment, on the outcome in a novel way. We
illustrate the models by fitting them to a large data set from a
clinical trial of the drug Naltrexone. The~HMM, in particular, fits
this data well and also contains unique features that allow for useful
clinical interpretations of alcohol consumption behavior.
\end{abstract}

%

\end{frontmatter}
%

\section{Introduction}
A major goal in alcoholism research is to develop models that can
accurately describe the drinking behavior of individuals with Alcohol
Use Disorders (AUDs: alcohol abuse or alcohol dependence), so as to
better understand common patterns among them. Clinical trials of
treatments for AUDs are run with the closely related goal of
understanding the relationship between a subject's alcohol consumption
and treatment variables, as well as certain background variables, like
age and sex, and other time-varying variables, like a daily measure of
a subject's desire to drink and/or stress level [McKay et al. (\citeyear{McFrPaLy2006})].
It is of particular interest to understand the relationship between
these variables and relapses into AUD, as opposed to ordinary alcohol
use. In summary, alcohol research is an area in which statistical
models are needed to provide rich descriptions of alcohol consumption
as a stochastic process in such a way that we can make inferences about
the relationship between exogenous variables and drinking. The~main
focus of this paper is developing models that are flexible enough to
describe a wide variety of drinking behaviors, and are parsimonious
enough to allow for clinically interpretable inferences.

We propose nonhomogeneous, hierarchical Bayesian Markov models and
hidden Markov models (HMMs) with random effects for AUDs treatment
trial data. Trial subjects' drinking behavior over time is often
characterized by flat stretches of consistent alcohol consumption,
interspersed with bursts of unusual consumption. Markov models and HMMs
are well-suited to model flat stretches and bursts, enabling them to
make good predictions of future drinking behavior (compared to other
models). We incorporate covariate and treatment effects into the
transition matrices of the Markov models and HMMs, which allows a rich
description of the evolution of a subject's drinking behavior over
time. The~inclusion of random effects into these models allows them to
account for unobserved heterogeneity on the individual level.
Furthermore, the HMM, which this paper focuses on more than the Markov
model, contains features that lend themselves to useful clinical
interpretations, and provides a quantitative model of a widely used
theoretical model for relapses known among alcohol researchers as the
cognitive-behavioral model.

As an illustration, we fit these models to a large ($N=240$
individuals, $T=168$ time points) data set from a clinical trial of the
drug Naltrexone that was conducted at the Center for Studies of
Addictions (CSA) at the University of Pennsylvania. We fit the models
using MCMC methods, and examine their fits with respect to various
clinically important statistics using posterior predictive checks.
These statistics include such things as the time until a subject
consumes his or her first drink, the proportion of days in which a
subject drinks, and various measures of serial dependence in the data.
We also estimate the treatment effect in a novel way by estimating its
effect on transitions between latent states.

This paper builds on recent work in using HMMs for large longitudinal
data sets. Altman (\citeyear{Al2007}) unifies previous work on HMMs for multiple
sequences [Humphreys (\citeyear{Hu1998}), Seltman (\citeyear{Se2002})] by defining a new class of
models called Mixed Hidden Markov Models (MHMMs), and describes a
number of methods for fitting these models, which can include random
effects in both the hidden and conditional components of the
likelihood. The~models we fit are similar to those described in Altman
(\citeyear{Al2007}), but our data sets are substantially larger, and we choose a
different method for fitting the models, using a Bayesian framework
outlined by Scott (\citeyear{Sc2002}). This work also provides another example of
using HMMs for health data, as was done by Scott, James and Sugar (\citeyear{ScJaSu2005}) in a
clinical trial setting with many fewer time points than the one
examined here.

The~rest of the paper is organized as follows: Section \ref{sec2} introduces the
the clinical trial that generated the data to which we fit our models,
and discusses existing models and the motivation for new models.
Section \ref{sec3} introduces the HMM and Markov model that we fit to the data.
Section \ref{sec4} describes the MCMC algorithm we used to fit the models and
some model selection criteria we considered when comparing the fits of
each model. Section \ref{sec5} provides a basic summary of the parameter
estimates of the models. Section \ref{sec6} discusses the effects of the
treatment and other covariates in terms of the transition probabilities
and the stationary distribution of the HMM. Section \ref{sec7} discusses the fit
of the HMM via posterior predictive checks of some common
alcohol-related statistics, and compares the fit of the HMM to that of
the Markov model in terms of the serial dependence structure of the
data. Section \ref{sec8} discusses how the HMM can be used to provide a new
definition of a relapse, which is a well known problem in the alcohol
literature. Last, Section \ref{sec9} provides a summary.

\section{Background and data}\label{sec2}
Before introducing the models we fit, we discuss the features of the
AUD treatment clinical trial from which we collected our data, some
existing models for alcohol data and the motivation for using Markov
models and HMMs.

\subsection{Data description}\label{sec2.1}
The~data set to which we fit our models is from a clinical trial of a
drug called Naltrexone conducted at the CSA at the University of
Pennsylvania. It contains daily observations from $N = 240$ subjects
over $T = 168$ days (24 weeks, or about 6 months). The~subjects were
volunteers who had been diagnosed with an AUD, and were therefore prone
to greater than average alcohol consumption and more erratic drinking
behavior than individuals without an AUD diagnosis. The~subjects
self-reported their alcohol consumption, making measurement error a
likely possibility. Immediately prior to the trial, the subjects were
required to abstain from drinking for at least three days as a
detoxification measure. Subjects recorded their daily consumption in
terms of a standard scale in alcohol studies, where ``one standard
drink'' represents 1.5 oz. of hard liquor, or 5~oz. of 12\%
alcohol/volume wine, or 12 oz. of (domestic US) beer. In order to
reduce the influence of outliers and possible measurement error, it is
common [Anton et al. (\citeyear{Anetal2006})] to code the number of drinks consumed as
an ordinal variable with three levels, corresponding to no drinking (0
drinks), light drinking (1--4 drinks) and heavy drinking (four or more
standard drinks for women, five or more standard drinks for men). A
complete description of the trial is available in Oslin et al. (\citeyear{OsLyPeirkt2008}).

\begin{figure}

\includegraphics{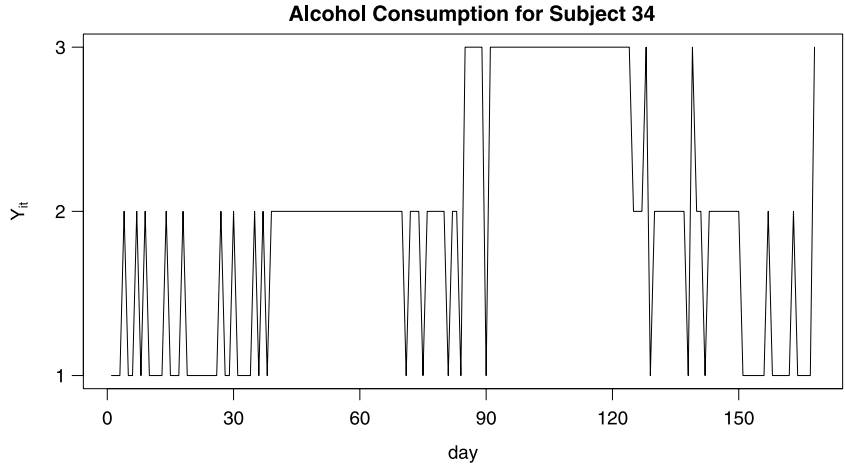}

\caption{The~time series plot of alcohol consumption for Subject 34.
The~vertical axis has tickmarks at 1 (0 drinks), 2 (1--4 drinks) and 3
(5 or more drinks).}\label{fig:ts34}
\end{figure}

\begin{figure}[b]

\includegraphics{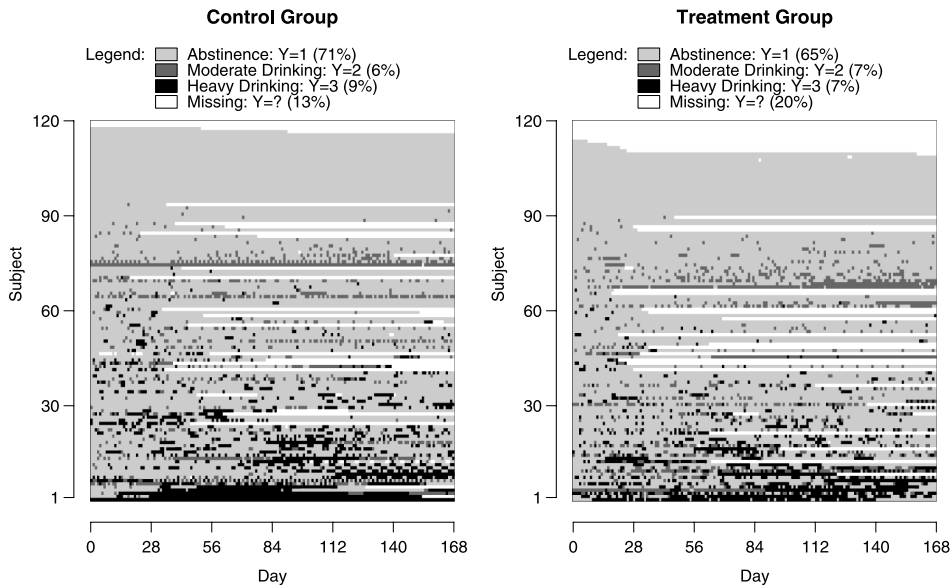}

\caption{A representation of the observed alcohol consumption time
series for all $N=240$ subjects in the clinical trial. They are
approximately sorted from top to bottom (within their treatment group)
in order of increasing alcohol consumption.}
\label{fig_data_picture}
\end{figure}

A sample series for one of the subjects is plotted in Figure~\ref
{fig:ts34}, and the entire collection of 240 time series are
represented in Figure~\ref{fig_data_picture}. The~overall frequencies
of the dependent variable are the following: 0 drinks ($Y=1$, 68\%),
1--4 drinks (\mbox{$Y=2$}, 7\%), $4/5$ or more drinks for women/men ($Y=3$, 8\%),
with 17\% of observations missing.

We include 4 covariates in the models we fit:
\begin{enumerate}
\item Treatment. Half the subjects were randomly assigned to receive
the drug Naltrexone, and the other half were assigned a placebo.

\item Sex. There were 175 men in the study, and 65 women.

\item Previous drinking behavior. Each study participant was asked to
estimate the proportion of days prior to the study during which they
drank, and during which they drank heavily. We used these two
variables, denoted $D_{(i,\mathrm{drink})}$ and $D_{(i,\mathrm{heavy})}$,
to calculate the proportion of days prior to the trial during which
each subject drank moderately, denoted $D_{(i,\mathrm{moderate})} =
D_{(i,\mathrm{drink})} - D_{(i,\mathrm{heavy})}$, and then created a
``previous drinking behavior'' index, $Z_i = D_{(i,\mathrm{moderate})} +
2D_{(i,\mathrm{heavy})} \in[0,2]$. This variable is a univariate
summary of a subject's previous drinking behavior which is
approximately unimodal.

\item Time. We measure time in days from $1,2,\ldots ,T=168$. Subjects did
not necessarily participate during the same stretch of actual calendar
days, so this variable is only a measure of the relative time within
the clinical trial.
\end{enumerate}

Prior to fitting the models, we scale these 4 input variables to have a
mean of zero and a standard deviation of $1/2$, so that the regression
coefficients are comparable on the same scale. The~ordinal drink counts
and covariates for all 240 subjects are available as supplementary
material in Shirley et al. (\citeyear{ShSmLyMaOs2009}).

\subsection{Existing models}\label{sec2.2}
Current models for alcohol data range from relatively simple models
that compare univariate summary statistics among different treatment
groups to more complex failure time models, linear mixed models and
latent transition models. Wang et al. (\citeyear{WaWiMcNe2002}) provide a good review and
a discussion of why very simple summary statistics are not adequate for
modeling drinking outcomes over time. Some examples of such summary
statistics are average number of drinks per day and percentage of days
abstinent. Average drinks per day, for example, might not differentiate
between one subject who drinks moderately every day of the week and
another subject who abstains from drinking during the week, and then
binges on the weekend. Percentage of days abstinent, on the other hand,
does not differentiate between light and heavy drinking, and does not
provide information about the pattern of abstinence across time.

Wang et al. suggest the use of multivariate failure time models for
alcohol data. These models can be used to predict the amount of time
before the next relapse. One shortcoming of such models is that they
require a strict definition of a relapse in order to define an
``event.'' Although alcohol researchers agree on a conceptual definition
of a relapse as a failure to maintain desired behavior change, they do
not agree on an operational definition of relapse [Maisto et al.
(\citeyear{MaPoCoLyMa2003})]. For example, do two days of heavy drinking separated by a
single day of abstinence constitute two relapses or just one? Does one
isolated day of heavy drinking constitute a relapse or just a ``lapse''?
Any inference drawn from a multivariate failure time model is
conditional on the definition of a relapse. Maisto et al. (\citeyear{MaPoCoLyMa2003}) show
that the choice of relapse definition makes a meaningful difference in
estimates of relapse rates and time to first relapse.

McKay et al. (\citeyear{McFrPaLy2006}) review GEE models for repeated measures alcohol
data. Although GEE models are useful in many ways, they are best suited
to estimate the population mean of drinking given a set of covariates,
rather than to make predictions for an individual given the
individual's covariates and drinking history. Latent transition models,
which are closely connected to HMMs, have been proposed in the
literature on addictive behavior [e.g., Velicer, Martin and Collins (\citeyear{VeMaCo1996})]. These
models typically require a multivariate response vector at each time
point, and are fit for data with a small number of time points, whereas
our data have a univariate response at each time point and a long time
series for each subject.

Mixed effects models are in principle well suited to making individual
predictions [McKay et al. (\citeyear{McFrPaLy2006})], but a rich structure for the serial
correlation must be considered to accommodate the flat stretches and
bursts nature of alcohol data, something which has not been considered
thus far in the alcohol research literature. This paper introduces two
such models---a mixed effects HMM and a mixed effects Markov model for
alcohol data.

\subsection{Motivation for a Markov model or HMM}\label{sec2.3}
One of the important goals of this research is to fit a model to
alcohol data that can provide a rich description of the changes in a
subject's drinking behavior through time. A first-order Markov (chain)
model is a suitable starting point in the search for such a model. It
models the observed data at a high resolution---daily observations---as
opposed to some lower-dimensional summary statistic of the data, such
as the time until the first relapse. By modeling the observations
themselves, we can infer the distribution of any lower-dimensional
summary statistic we wish after fitting the model. The~serial
dependence structure of a Markov model, however, is such that it may
not be suitable for data that exhibit long-term dependence.

The~HMM is an attractive alternative to the first-order Markov model
because its serial dependence structure allows for longer-term
dependence. As we have stated previously, drinking behavior among
individuals following their AUD treatment is erratic, commonly
exhibiting flat stretches and bursts, creating long-term dependence.
Longitudinal data of this form suggest that a \textit
{parameter-driven} model might fit the data better than an \textit
{observation-driven} model. Cox (\citeyear{Co1981}) originally coined the term
parameter-driven for models in which an underlying, potentially
changing ``parameter process'' determines the distribution of observed
outcomes through time, such that the observed outcomes depend on each
other only through the underlying parameter process. Observation-driven
models, on the other hand, model observed outcomes directly as
functions of previously observed outcomes. State-space models and HMM's
(which are a subset of state-space models) are widely used examples of
parameter-driven models.

A second, very important motivation for the HMM for alcohol data is its
potential clinical interpretation: it is a quantitative model that
closely corresponds to a well-developed theoretical model of relapse
known as the cognitive-behavioral model of relapse. This theory
suggests that the cause of an alcohol relapse is two-fold: First, the
subject must be in a mental and/or physical condition that makes him or
her vulnerable to engaging in an undesired level of drinking. In other words,
if the subject were given an opportunity to drink alcohol, he or she
would be unable to mount an effective coping response, that is, a way
to avoid drinking above some desired level (usually abstinence or a
moderate level of drinking). A subject's ability to cope can depend on
time-varying cognitive factors, such as expectations of the results of
drinking, time-varying noncognitive factors, such as level of stress,
desire to drink and social support, and background variables such as
age or sex. The~second condition that must be present for a relapse is
that the subject encounters a high-risk drinking situation, which is a
setting associated with prior heavy alcohol use. Common examples are a
negative mood, interpersonal problem(s) and an invitation to drink.
McKay et al. (\citeyear{McFrPaLy2006}) and Marlatt and Gordon (\citeyear{MaGo1985}) contain more detailed
discussions of the traditional cognitive-behavioral model of relapse,
and Witkiewitz and Marlatt (\citeyear{WiMa2004}) update Marlatt and Gordon (\citeyear{MaGo1985}).

The~HMM connects to the cognitive-behavioral model of relapse in the
following way: The~hidden states in the HMM correspond to the
vulnerability of the subject to drinking if faced with a high risk
situation, and the number of drinks consumed on a given day is affected
by the subject's vulnerability to drinking on the given day (the hidden
state) and whether the subject faced a high-risk drinking situation on
that day. The~conditional distributions of the observations given the
hidden states model the prevalence of high-risk drinking situations for
each hidden state. Interpreting the HMM this way suggests that relapse
is not necessarily an observable state---it is a hidden state that
presumably leads to heavy drinking with higher probability than other
states, but not with certainty. Via this correspondence to the
cognitive-behavioral model of relapse, the HMM has potential to make an
important contribution to the alcohol literature in its ability to
provide quantitative insight into a popular qualitative model for relapse.

\section{The~models}\label{sec3}
The~data we wish to model consist of categorical observations $Y_{it}
\in(1,2,\ldots ,M)$, for subjects $i = 1, \ldots , N$, and time points $t =
1,\ldots , n_i$ (where $n_i=168$ for all subjects in our data set). We also
observe the $p \times  1$ covariate vectors $\mathbf{X}_{it} = (X_{it1}, X_{it2},
\ldots , X_{itp})^{\prime}$ for each subject at each time point. We consider two
types of models in this paper: (1) a hidden Markov model, and (2)~a~Markov model.

\subsection{Hidden Markov model}\label{sec3.1}
The~HMM we fit consists of a mixed effects multinomial logit model for
each row of the hidden state transition matrix, and multinomial models
for (1) the conditional distributions of the observations given the
hidden states, hereafter called the conditional distributions of the
HMM, and (2) the initial hidden state distribution. The~model includes
an unobserved, discrete hidden state, $H_{it} \in(1,2,\ldots ,S)$ for each
individual $i$ at each time point $t$, and we denote an HMM with $S$
hidden states as an HMM($S$).

The~hidden state transition matrix is an $S\times S$ matrix, in which
each row has probabilities that are modeled using a mixed effects
multinomial logit model:
%
\begin{equation}\label{hmm_tprobs}
p\bigl(H_{(i,t+1)} = s \vert  H_{it} = r, \mathbf{X}_{it}, \bolds{\alpha
}_{ir}, \bolds{\beta}_r\bigr)
= \frac{\exp(\alpha_{irs} + \mathbf{X}_{it}^{\prime}\bolds{\beta
}_{rs})}{1 + \sum
_{k=2}^S\exp(\alpha_{irk} + \mathbf{X}_{it}^{\prime}\bolds{\beta}_{rk})}
\end{equation}
for $r=1,\ldots ,S$, $s = 2,\ldots ,S$, $i=1,\ldots ,N$, and $t=1,\ldots ,T-1$. For
$s=1$, the numerator is set equal to 1, making the first category the
baseline category. The~random intercepts are modeled using normal
distributions, where $\alpha_{irs} \sim N(\mu_{rs},\sigma_{rs}^2),$
for $i=1,\ldots ,N$, $r=1,\ldots ,S$, and $s=2,\ldots ,S$.

The~inclusion of $S(S-1)$ random intercepts for each subject means that
each subject's transition matrix is fully flexible to reproduce every
possible set of cell frequencies for that subject. The~regression
effects, on the other hand, are ``fixed'' effects that are constant
across all subjects. They may, however, be time-varying.

The~conditional distributions of the HMM are multinomial distributions, where
%
\begin{equation}\label{model_hmm_cond}
p(Y_{it} = m | H_{it} = s, \mathbf{p}_s) = p_{sm}
\end{equation}
for $m=1,\ldots ,M$ and $s=1,\ldots ,S$, where $\mathbf{p}_s = (p_{s1}, p_{s2},
\ldots , p_{sM})$ is a vector of multinomial probabilities, for $s =
1,\ldots ,S$. Last, the initial hidden state distribution for the HMM is a
multinomial distribution with probability vector $\bolds{\pi}= (\pi
_1,\ldots ,\pi_S)$. For the hidden Markov model, the collection of all
parameters is $\bolds{\theta}= (\bolds{\alpha}, \bolds{\beta},
\bolds{\mu}, \bolds{\sigma}, \bolds{\pi},
\mathbf{P})$.

\subsection{Markov model}\label{sec3.2}
The~other model we consider is a first-order Markov model, in which the
set of probabilities in each row of the transition matrix is defined
according to the same model described in Section \ref{sec3.1}, a
mixed effects multinomial logit model. The~outcome for individual $i$
at time $t+1$, $Y_{(i,t+1)}$, depends on the previous observation for
that individual, $Y_{it}$, in the following way:
%
\begin{equation}\label{model_markov}
p\bigl(Y_{(i,t+1)} = m \vert  Y_{it}=j, \mathbf{X}_{it}, \bolds{\alpha
}_{ij}, \bolds{\beta}_j\bigr) =
\frac{\exp(\alpha_{ijm} + \mathbf{X}_{it}^{\prime}\bolds{\beta
}_{jm})}{1 + \sum
_{k=2}^M\exp(\alpha_{ijk} + \mathbf{X}_{it}^{\prime}\bolds{\beta}_{jk})}
\end{equation}
for $j=1,\ldots ,M$, $m = 2,\ldots ,M$, $i=1,\ldots ,N$, and $t=1,\ldots ,T-1$. As
before, for $m=1$, the numerator equals 1. The~random intercepts are
given normal distributions, where $\alpha_{ijm} \sim N(\mu
_{jm},\sigma_{jm}^2),$ for $i=1,\ldots ,N$, $j=1,\ldots ,M$, and $m=2,\ldots ,M$.

The~initial values, $Y_{i1}$, for $i=1,\ldots ,N$, are modeled as
i.i.d.
draws from a multinomial distribution with probability vector $\bolds
{\pi}=
(\pi_1,\ldots ,\pi_M)$. For the Markov model, the collection of all
parameters is $\bolds{\theta}= (\bolds{\alpha}, \bolds{\beta},
\bolds{\mu}, \bolds{\sigma}, \bolds{\pi})$.

\subsection{Missing data}\label{sec3.3}

Following Oslin et al. (\citeyear{OsLyPeirkt2008})'s analysis of the trial we are
considering, we will assume that the missing data is missing at random,
that is, the missingness mechanism is independent of the missing data
given the observed data and covariates. For the HMM, we make the
additional assumption that the missingness mechanism is independent of
the hidden states given the observed data and covariates, in order to
do inference on the hidden states as well as the model parameters
[Gelman et al. (\citeyear{GeCaStRu2004}), Chapter 21]. A sensitivity analysis for this
assumption is a valuable topic for future research, but outside the
scope of this paper. There is a considerable literature on missing data
in longitudinal addiction studies, including Albert (\citeyear{Al2000}, \citeyear{Al2002}) and
Longford et al. (\citeyear{LoElHaWa2000}), which might guide one's choices regarding the
design of a sensitivity analysis.

\subsection{Discussion of models}\label{sec3.4}

The~first-order Markov model is nested within an HMM with $M$ or more
hidden states. Also, an HMM with $S$ hidden states is nested within an
HMM with more than $S$ hidden states. An HMM is a mixture model, whose
mixing distribution is a first-order Markov chain. It has been applied
to problems in numerous fields [for lists of examples, see MacDonald
and Zucchini (\citeyear{MaZu1997}), Scott (\citeyear{Sc2002}) or Cappe, Moulines and Ryden (\citeyear{CaMoRy2005})], but not
previously to alcohol consumption time series data.

\section{Model fits and comparison}\label{sec4}

\subsection{Model fits}\label{section_modelfits}

We fit the HMM(3) and the Markov model using\break Metropolis-within-Gibbs MCMC algorithms whose details are described in the \hyperref[section_appendix]{Appendix}.
 We ran three chains for each model and discarded
the first 10,000 iterations of each chain as burn-in. We used the next
10,000 iterations as our posterior sample and assessed convergence
using the potential scale reduction factor [Gelman et al. (\citeyear{GeCaStRu2004})], and
found that $\hat{R} < 1.02$ for all parameters in the models. Visual
inspection of trace plots provide further evidence that the posterior
samples converged to their stationary distributions. The~autocorrelation of the posterior samples of $\bolds{\beta}$ were the highest
of all the parameters, such that total effective sample size from three
chains of 10,000 iterations each was between 200 and 300. The~transition probabilities for each subject and time point, on the other
hand, had much lower autocorrelations, such that for a given transition
probability, three chains of 10,000 iterations each yielded a total
effective sample size of about 5000 approximately independent
posterior samples. Since transition probabilities are the most
important quantity of interest in our analysis, the Monte Carlo error
associated with our estimated treatment effects is small.

All the parameters were given noninformative or weakly informative
priors, which are described in the \hyperref[section_appendix]{Appendix}. The~algorithm for both models consisted of sampling some of the parameters
using univariate random-walk Metropolis steps, and others from their
full conditional distributions. When fitting the Markov model, if a
subject's data were missing for $t_{\mathrm{mis}}$ steps, we computed
the $(t_{\mathrm{mis}}+1)$-step transition probability to calculate the
likelihood of the current parameters given the data, and alternated
between sampling $\mathbf{Y}_{\mathrm{mis}} \sim p(\mathbf{Y}_{\mathrm
{mis}} \vert \mathbf{Y}
_{\mathrm{obs}},\mathbf{X},\bolds{\theta})$ and $\bolds{\theta}\sim
p(\bolds{\theta}\vert \mathbf{Y}
_{\mathrm{obs}},\mathbf{Y}_{\mathrm{mis}},\mathbf{X})$. Likewise for
the HMM, we
alternated sampling $\mathbf{H}\sim p(\mathbf{H}\vert \mathbf
{Y}_{\mathrm{obs}}, \mathbf{X}
,\bolds{\theta})$ and $\bolds{\theta}\sim p(\bolds{\theta}\vert
\mathbf{Y}_{\mathrm{obs}},\mathbf{H}
,\mathbf{X})$. We sampled $\mathbf{H}$ using a forward-backward
recursion as
described in Scott (\citeyear{Sc2002}), substituting $(t_{\mathrm{mis}}+1)$-step
transition probabilities between hidden states when there was a string
of $t_{\mathrm{mis}}$ missing observations for a given subject.

The~hidden states of an HMM have no {a priori} ordering or
interpretation, and as a result, their labels can ``switch'' during the
process of fitting the model from one iteration of the MCMC algorithm
to the next, without affecting the likelihood [Scott (\citeyear{Sc2002})]. We
experienced label switching when the starting points for the MCMC
algorithm were extremely widely dispersed, but when we used more
narrowly dispersed starting points, each chain converged to the same
mode of the $S!$ symmetric modes of the posterior distribution. We
discuss our procedure for selecting starting points for the parameters
in the MCMC algorithms in the \hyperref[section_appendix]{Appendix}.

\subsection{Model comparison}\label{sec4.2}

We present the DIC of the two models in this section, but we ultimately
don't rely on it as a formal model selection criterion. Instead we use
posterior predictive checks to compare the two models, and combine this
analysis with existing scientific background knowledge to motivate our
model choice, which suggests that the HMM(3) is more promising.

The~Markov model converged to a slightly lower mean deviance than the
HMM(3), where the deviance is defined as $D(\mathbf{Y}, \bolds{\theta
}) = -2\log
{p(\mathbf{Y}\vert \bolds{\theta})}$. Table~\ref{tab_DIC} contains a
summary of the
deviance for both models, where $\bar{D}(\mathbf{Y}, \bolds{\theta
})$ is the mean
deviance across posterior samples of $\bolds{\theta}$, and $D(\mathbf
{Y}, \bar
{\bolds{\theta}})$ is the deviance evaluated at the posterior mean,
$\bar
{\bolds{\theta}}$. For the Markov model, deviance calculations were done
using the parameter vector $\bolds{\theta}= (\bolds{\alpha}, \bolds
{\beta}, \bolds{\pi})$,
whereas for the HMM(3), the parameter vector $\bolds{\theta}= (\bolds
{\alpha},
\bolds{\beta}, \bolds{\pi}, \mathbf{P})$. That is, for both models,
we compute the
probability of the data given the parameters in the likelihood
equation, and don't factor in the prior probabilities of the random
effects, $p(\bolds{\alpha}\vert \bolds{\mu}, \bolds{\sigma})$. The~Markov model has a
higher estimated effective number of parameters, $pD$, and also has a
higher DIC, which is an estimate of a model's out of sample prediction
error [Spiegelhalter et al. (\citeyear{SpBeCaVa2002})]. The~estimates of $pD$ and DIC had
low Monte Carlo estimation error (approximately ${\pm}$5), as estimated by
recomputing them every 1000 iterations after burn-in.

\begin{table}[b]
\caption{Summary of the DIC of the models}\label{tab_DIC}
\begin{tabular*}{\textwidth}{@{\extracolsep{\fill}}lcccc@{}}
\hline
\textbf{Model} & $\bolds{\bar{D}(\mathbf{Y}, \bolds{\theta})}$ & $\bolds{D(\mathbf{Y},\bar{\bolds{\theta}})}$ & $\bolds{pD}$ &\textbf{DIC} \\
\hline
Markov & 19,871 & 19,183 & 688 & 20,560 \\
HMM(3) & 19,902 & 19,303 & 599 & 20,500 \\
\hline
\end{tabular*}
\end{table}

Model selection criteria such as BIC and DIC, however, are well known
to be problematic for complicated hierarchical models, mixture models
such as an HMM and models with parameters that have non-normal
posterior distributions [Spiegelhalter et al. (\citeyear{SpBeCaVa2002}), Scott (\citeyear{Sc2002})].
For this reason, we choose not to select a model [between the Markov
model and the HMM(3)] based on formal criteria such as these. Instead,
we examine the fit of each model using posterior predictive checks and
highlight each of their advantages and disadvantages. We ultimately
focus most of our attention on the HMM(3) because of its potential for
clinical interpretability in terms of the cognitive behavioral model of
relapse (Section~\ref{sec2.3}), which suggests that a relapse
is not equivalent to a heavy drinking episode, but rather, it may be
better modeled as an unobserved state of health.

\section{Interpreting the model fits}\label{sec5}
We begin by looking at the parameter estimates of the Markov model and HMM(3).

\subsection{Conditional distributions of the HMM}\label{section_conditional}

The~hidden states of the HMM(3) can be characterized by their
conditional distributions, which are multinomial distributions with
three cell probabilities, $p_{sm} = p(Y_{it}=m \vert  H_{it}=s)$ for
outcomes $m=1,2,3$ and hidden states $s=1,2,3$. The~posterior
distributions of $p_{sm}$ are summarized in Table~\ref{tab_cond_dists}. Each conditional distribution puts most of its mass
on a single outcome, and we therefore label the hidden states according
to which outcome has the largest probability under their associated
conditional distributions: The~first hidden state is labeled
``Abstinence'' (``A''), the second hidden state is labeled ``Moderate
Drinking'' (``M''), and the third hidden state is labeled ``Heavy
Drinking'' (``H'').
The~Moderate and Heavy Drinking states each leave a small, but
significant (as indicated by their 95\% posterior intervals excluding
zero) probability of a different outcome from moderate ($Y=2$) and
heavy ($Y=3$) drinking, respectively.

\begin{table}[b]
\caption{Conditional Distributions of the HMM(3)}\label{tab_cond_dists}
\begin{tabular*}{\textwidth}{@{\extracolsep{\fill}}lccc@{}}
\hline
& \textbf{Posterior mean} & \textbf{Posterior stand. errors} & \textbf{State label} \\
\hline
$\hat{\mathbf{p}}_1$ & (0.997, 0.003, 0.000) & (0.001, 0.001, 0.000) &``Abstinence'' \\
$\hat{\mathbf{p}}_2$ & (0.026, 0.956, 0.018) & (0.012, 0.012, 0.005) &``Moderate Drinking'' \\
$\hat{\mathbf{p}}_3$ & (0.031, 0.004, 0.966) & (0.006, 0.002, 0.007) &``Heavy Drinking'' \\
\hline
\end{tabular*}
\end{table}

\subsection{Transition probabilities}\label{sec5.2}
The~posterior distributions of the parameters associated with the
transition matrices for the Markov model and the HMM(3) are similar.
This is not surprising, considering that if $p_{11}=p_{22}=p_{33}=1$,
then the HMM(3) is equivalent to the Markov model, and from Table~\ref
{tab_cond_dists} we see that the estimates of $p_{11}, p_{22}$ and
$p_{33}$ are almost equal to 1. We will not discuss the posterior
distributions of these parameter estimates here, though, because the
form of the multinomial logit model for each row of the transition
matrices makes them somewhat difficult to interpret. Instead, we
present the estimated mean transition matrices for both models:
%
\begin{equation}\label{trans_matrices}
\hspace*{12pt} \hat{Q}_{\mathrm{Markov}} = \pmatrix{
0.97 & 0.02 & 0.01\vspace*{2pt}\cr
0.75 & 0.21 & 0.04\vspace*{2pt}\cr
0.45 & 0.03 & 0.52},
\qquad
\hat{Q}_{\mathrm{HMM(3)}} = \pmatrix{
0.98 & 0.01 & 0.01\vspace*{2pt}\cr
0.72 & 0.26 & 0.02\vspace*{2pt}\cr
0.41 & 0.02 & 0.57}.\hspace*{-12pt}
\end{equation}
These are the posterior mean transition matrices for the Markov model
and HMM(3) for a subject with average random intercepts, and the
average value for each covariate: $\mathbf{X}_i=(0,0,0,0)^{\prime}$. Formally,
\[
\hat{Q}_{\mathrm{Model}}(j,m) = \frac{1}{G}\sum_{g=1}^{G}
\biggl(\frac{\exp{(\mu_{jm}^{(g)})}}{1+\exp{(\mu_{j2}^{(g)})}+\exp{(\mu
_{j3}^{(g)})}}\biggr)
\]
for $j=1,\ldots ,3$, $m=2,3$, and posterior samples $g=1,\ldots ,G$, where for
$m=1$, the numerator is replaced by the value 1. Setting the time
variable equal to zero corresponds to calculating the mean transition
matrix on day 84 of the 168-day clinical trial. Note that Sex and
Treatment are binary variables, which means that no individual subject
has the exact design matrix used to calculate the transition
probabilities in Equation (\ref{trans_matrices}). Later in
Section~\ref{sec6} we will more carefully analyze the
transition matrices bearing this in mind, but for now we are only
interested in rough estimates of the transition probabilities.

The~estimates of the mean transition probabilities are very similar
between the two models. One interesting feature is that, on average,
for both models, the most persistent state is Abstinence, the second
most persistent state is Heavy Drinking, and the least persistent state
is Moderate Drinking. The~Moderate Drinking and Heavy Drinking states
don't communicate with each other with high probability under either
model (for a person with average covariates). We fully analyze the
covariate effects on the transition matrix in Section~\ref{sec6}.

\subsection{Initial state probabilities}\label{sec5.3}
The~posterior distributions of the initial state probability vectors
for the two models were nearly identical, where their posterior means
were $\hat{\bolds{\pi}} = (93.6\%,3.4\%,3.0\%)$ for both models, and the
posterior standard deviation was about 1.5\% for all three elements of
$\bolds{\pi}$ (for both models).

\section{The~treatment effect}\label{sec6}
To estimate the effects of covariates, including the treatment
variable, on the transition probabilities of the hidden state
transition matrix for the HMM(3) (where a similar analysis could be
done for the Markov model if desired), we perform average predictive
comparisons as described in Gelman and Hill (\citeyear{GeHi2007}), pages 466--473. The~goal
is to compare the value of some quantity of interest, such as a
transition probability, calculated at two different values of an input
variable of interest, such as Treatment, while controlling for the
effects of the other input variables. Simply setting the values of the
other inputs to their averages is one way to control for their effects,
but doing so can be problematic when the averages of the other inputs
are not representative of their distributions, as is the case for
binary inputs, for example. An average predictive comparison avoids
this problem by calculating the quantity of interest at two different
values of the input of interest while holding the other inputs at their
observed values, and averaging the difference of the two quantities
across all the observations.

In the context of the HMM(3), if we denote the input variable of
interest by $U$, and consider high and low values of this variable,
$u^{\mathrm{(hi)}}$ and $u^{\mathrm{(lo)}}$, and denote the other input
variables for individual $i$ at time $t$ by $\mathbf{V}_{it}$, then we can
compute the average difference in transition probabilities calculated
at the two values of $U$ as follows:
%
\begin{eqnarray}\label{eqn_apc}
B_{jm}(U) &=& \frac{1}{N(T-1)}\nonumber\\
&&{}\times \sum_{i=1}^N \sum_{t=1}^{T-1} \bigl[
\mathrm{P}\bigl(H_{(i,t+1)}=m \vert  H_{it}=j, \bolds{\theta}, u_{it}=u^\mathrm
{(hi)},\mathbf{v}_{it}\bigr)\\
&&\hspace*{47pt}{} - \mathrm{P}\bigl(H_{(i,t+1)}=m \vert  H_{it}=j, \bolds
{\theta},
u_{it}=u^\mathrm{(lo)},\mathbf{v}_{it}\bigr)\bigr] \nonumber
\end{eqnarray}
for $j,m=1,\ldots ,3$. We call $B_{jm}(U)$ the average transition
probability difference for input variable $U$ for transitions from
hidden state $j$ to hidden state $m$.

\begin{figure}

\includegraphics{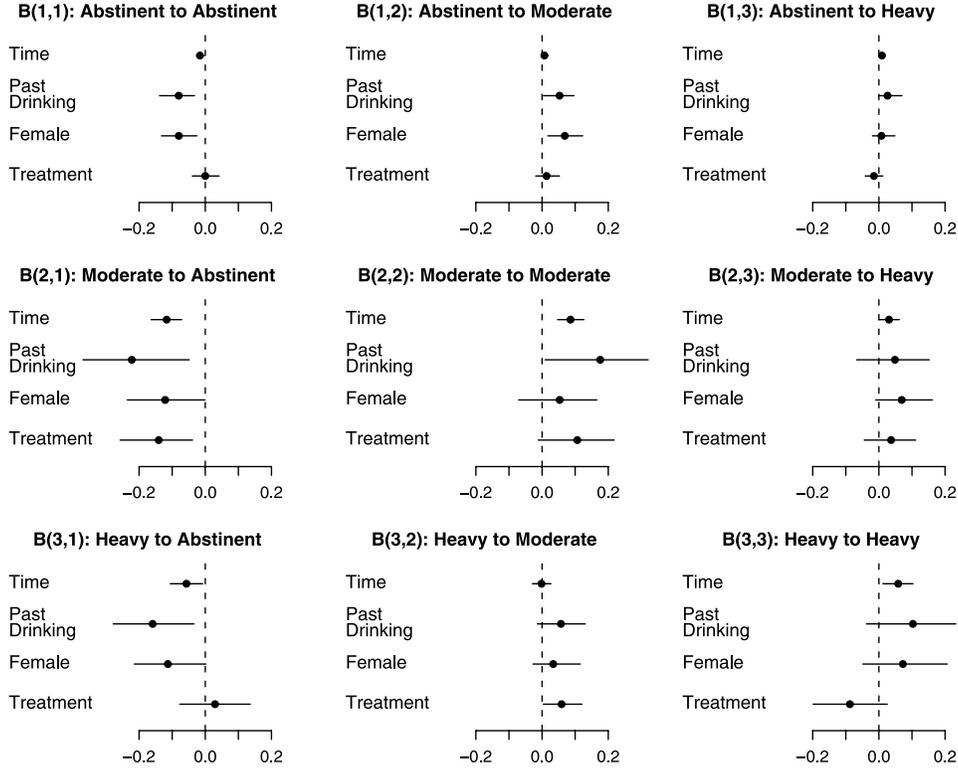}

\caption{95\% posterior intervals of the average transition
probability differences for each of the four input variables and for
each of the nine transition probabilities in the HMM(3). For the binary
input variables Treatment and Sex, we set the high values, $u^{\mathrm
{(hi)}}$, as Naltrexone and Female, and the low values, $u^{\mathrm
{(lo)}}$, as Control and Male, respectively, and for the continuous
input variables, Past Drinking and Time, we set the high and low values
to the mean $\pm$ 1 sd of their respective distributions.}
\label{fig_apc}
\end{figure}

Figure~\ref{fig_apc} displays 95\% posterior intervals for the average
transition probability differences for all four covariates that are
incorporated into the transition matrix of the HMM(3), for all $3^2=9$
possible transitions. The~treatment, Naltrexone, appears to have a
moderate negative effect on the probabilities of a transition from
Heavy-to-Heavy and Abstinent-to-Heavy, as is evidenced by the fact that
the 95\% posterior intervals for these average transition probability
differences lie mostly below zero [the intervals are $(-19.9\%, 2.5\%)$
and $(-4.1\%, 1.1\%)$, respectively]. Heavy-to-Heavy transitions occur,
on average, about 9\% less often for those using Naltrexone, and
Abstinent-to-Heavy transitions occur, on average, about 1.5\% less
often for those using Naltrexone. This is a desired outcome: These
intervals indicate that Naltrexone may be effective in improving
certain aspects of drinking behavior, because a reduction in these
transition probabilities would most likely result in a decrease in
heavy drinking, according to the estimated conditional distributions
associated with these hidden states. Conversely, Naltrexone appears to
have a significant positive effect on the probability of a transition
from Moderate-to-Heavy, occurring about 3.5\% more often on average for
Naltrexone-takers, which, on average,would lead to increased heavy drinking.

Other covariate effects visible in Figure~\ref{fig_apc} include the following:
\begin{itemize}
\item All else being equal, women are more likely than men to make
transitions into the Moderate and Heavy states.
\item Drinking heavily in the past is associated with higher transition
probabilities into the Moderate and Heavy states.
\item Transitions into the Moderate and Heavy states are more likely to
occur later in the clinical trial than earlier in the clinical trial.
\item The~95\% intervals for the average transition probability
differences are noticeably shorter for transitions out of the
Abstinence state, because that is where subjects are estimated to spend
the most time, and with more data comes more precise estimates.
\end{itemize}

\subsection{Covariate effects on the stationary distribution}\label{sec6.1}
To further summarize the effects of the covariates on drinking
behavior, we can compute average stationary probability differences
instead of average transition probability differences. This provides a
slightly more interpretable assessment of the treatment effect on
drinking behavior because the stationary distribution contains a
clinically important summary statistic: the expected percentages of
time a subject will spend in each hidden state, given that the HMM is
stationary. An effective treatment would reduce the expected amount of
time spent in the Heavy Drinking state.

\begin{figure}

\includegraphics{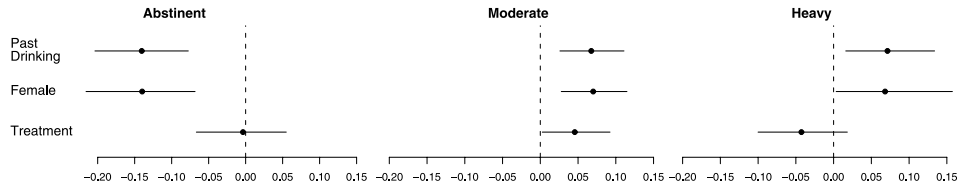}

\caption{95\% posterior intervals of average stationary probability
differences for the three covariates (other than time) incorporated
into the HMM(3) transition matrix.}
\label{fig_stat_dist}
\end{figure}

The~model we fit is a nonhomogenous (i.e., nonstationary) HMM, so, by
definition, it does not have a stationary distribution. We would still,
however, like to make inferences about a subject's long-term behavior
conditional on whether or not he or she took Naltrexone. One strategy
for estimating how much time will be spent in each state is through
posterior predictive simulations of the hidden states themselves.
Another strategy, which we discuss here, is to compute the stationary
distribution of the hidden state transition matrix on the last day of
the trial for all subjects. This can be interpreted as a way of
projecting drinking behavior into the long-term future without
extrapolating the effect of time beyond the clinical trial, by assuming
that each subject has reached a stationary pattern of behavior at the
end of the trial. Figure~\ref{fig_stat_dist} contains 95\% posterior
intervals of the average stationary probability differences for the
covariates Treatment, Sex and Past Drinking, where the HMM(3) is
assumed to be stationary beginning on the last day of the trial.

On average, the use of Naltrexone appears to decrease the amount of
time spent in the Heavy drinking state by about 5\%, and increase the
amount of time spent in the Moderate drinking state by the same amount.
Women and subjects who drank heavily prior to the trial also are
predicted to spend more time in the Heavy drinking state in the long
run. These estimates are consistent with the average transition
probability differences displayed in Figure~\ref{fig_apc}, and provide
a slightly more concise summary of the covariate effects.

\subsection{Heterogeneity among individuals}\label{sec6.2}
\begin{figure}

\includegraphics{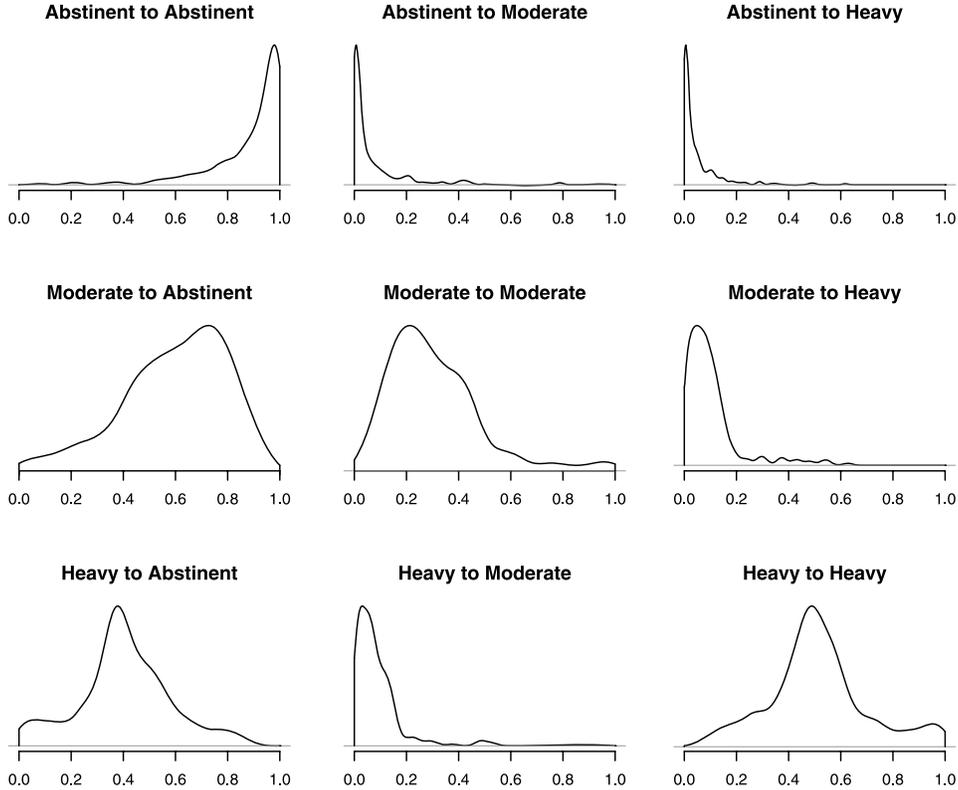}

\caption{Density estimates of the $3^2=9$ sets of posterior mean
transition probabilities in the HMM(3), where the posterior means are
calculated for each individual at the midpoint of the trial ($t=84$),
and each of the 9 the transition probability densities are therefore
estimated using $N=240$ data points.}
\label{fig_tpost}
\end{figure}

Another important aspect of the model fit is the degree of
heterogeneity that is present among individuals, which is modeled by
the random intercepts as well as the covariate effects embedded in the
HMM(3) hidden state transition matrix. To explore this, we calculate
the posterior mean of each of the $3^2=9$ transition probabilities for
each subject $i$ at a given time point $t$,
%
\begin{equation}
\hat{Q}_{it}(r,s) = \frac{1}{G}\sum_{g=1}^G \mathrm{P}\bigl(H_{(i,t+1)}=s
\vert  H_{it}=r, X_{it},\bolds{\alpha}^{(g)},\bolds{\beta
}^{(g)},\bolds{\pi}^{(g)}\bigr)
\end{equation}
for $i=1,\ldots ,N$, $t=84$ (the midpoint of the trial), $r,s=1,2,3$, and
posterior samples $g=1,\ldots ,G$. In Figure~\ref{fig_tpost} we plot
density estimates of each of the $3^2=9$ sets of $N$ transition
probability posterior means (measured at time $t=84$).

The~density estimates of these posterior mean transition probabilities
represent heterogeneity among individuals from two sources: the random
intercepts and the covariate effects. By comparing Figure~\ref
{fig_apc} to Figure~\ref{fig_tpost}, it is clear that the differences
in mean transition probabilities among individuals are larger than the
differences implied by covariate effects alone. The~largest average
transition probability differences, illustrated in Figure~\ref
{fig_apc}, were approximately ${\pm}$0.2, and were due to differences in
the subjects' previous drinking behavior. We see from Figure~\ref
{fig_tpost}, however, that the differences between transition
probabilities across individuals is often greater than 0.5, and can be
as large as 1, such that a particular transition is certain for one
individual, and has zero probability for another individual.
Specifically, transition probabilities are relatively similar across
subjects when the subjects are in the Abstinent state, but are highly
variable across subjects when the subjects are in the Moderate and
Heavy states. If a subject is in the Abstinent state, their probability
of remaining in the Abstinent state for an additional day is above 80\%
for most subjects, regardless of the values of his or her observed
covariates and random intercepts (which represent the effects of
unobserved covariates). On the other hand, if a subject is in the
Moderate or Heavy state, his or her next transition is highly
uncertain: depending on the individual, he or she may have a very high
or a very low probability of remaining in the same state for another
day. The~fact that the random intercepts are explaining more variation
in the estimated transition probabilities than any individual covariate
suggests that it might be useful for future studies to record as many
additional covariates as possible. Such a design might reduce the
amount of unexplained heterogeneity across individuals that must be
accounted for by the random intercepts. This result also highlights the
need for random effects in this particular model, as well as the need
for studying the inclusion of random effects into HMMs in general, as
they can have a profound effect on the fit of the model.

\section{Goodness of fit of the models}\label{sec7}

In this section we use posterior predictive checks to examine the fit
of the HMM(3) and the difference between the HMM(3) and the Markov model.

\subsection{Mean and variance of drinking days across subjects}\label{sec7.1}

As a first check of the goodness of fit of our model, we performed
posterior predictive checks [Gelman et al. (\citeyear{GeCaStRu2004})] of the mean and
variance of the number of moderate and heavy drinking days across all
subjects (a total of four statistics of interest). The~specific steps
we took are as follows: For each of these four statistics, we computed
its observed value from the data, and denoted this $t(\mathbf{Y}^\mathrm
{obs})$. Then, for posterior samples $g=1,\ldots ,G$, we
\begin{enumerate}
\item simulated $\bolds{\alpha}^\mathrm{rep} \vert \bolds{\mu}^{(g)},
\bolds{\sigma}
^{(g)}$, where $\bolds{\alpha}^\mathrm{rep}$ is a set of random
effects to
represent a new set of subjects from the population,
\item simulated $\mathbf{H}^\mathrm{rep} \vert \bolds{\alpha}^\mathrm
{rep}, \bolds{\beta}
^{(g)}, \bolds{\pi}^{(g)}, \mathbf{X}$,
\item simulated $\mathbf{Y}^\mathrm{rep} \vert \mathbf{H}^\mathrm{rep},
\mathbf{P}^{(g)}$, and
\item computed $t(\mathbf{Y}^\mathrm{rep})$,
\end{enumerate}
where the parameters superscripted by $(g)$ are taken from the
posterior samples obtained from the MCMC algorithm. By simulating new
random effects, $\bolds{\alpha}^\mathrm{rep}$, we are checking to see
that the
model adequately captures heterogeneity across subjects. We compared
the values of $t(\mathbf{Y}^\mathrm{obs})$ to the distributions of
$t(\mathbf{Y}
^\mathrm{rep})$ for the four statistics of interest, and found no
evidence of a lack of fit; that is, the observed values of $t(\mathbf{Y}
^\mathrm{obs})$ were not located in the extreme tails of the
distributions of $t(\mathbf{Y}^\mathrm{rep})$. Table~\ref{tab_pp} contains
summaries of these posterior predictive checks.

\begin{table}
\caption{Summaries of the posterior predictive checks of the mean and
variance of moderate and heavy drinking days. $s^2_\mathrm{mod}$ and
$s^2_\mathrm{heavy}$ are the variances of the number of moderate and
heavy drinking days across subjects, respectively, and $\bar{x}_\mathrm
{mod}$ and $\bar{x}_\mathrm{heavy}$ are the means of the number of
moderate and heavy drinking days across subjects, respectively. The~middle 5 columns contain the 2.5th,
25th,~50th,~75th and 97.5th quantiles of the
distributions of $t(\mathbf{Y}^\mathrm{rep})$, and the rightmost column
contains the observed values $t(\mathbf{Y}^\mathrm{obs})$, and the proportion
of the simulated statistics that were~less~than~these~values}\label{tab_pp}
\begin{tabular*}{\textwidth}{@{\extracolsep{\fill}}lcccccc@{}}
\hline
\textbf{Statistic} & \textbf{2.5\%} & \textbf{25\%} & \textbf{50\%} & \textbf{75\%} & \textbf{97.5\%} & \textbf{Observed (quantile)}\\
\hline
$s^2_\mathrm{mod}$ & 335.9 & 393.5 & 430.6 & 477.6 & 554.6 & 524.8 (0.92) \\[2pt]
$s^2_\mathrm{heavy}$ & 400.6 & 502.9 & 558.4 & 608.4 & 722.7 & 618.5 (0.78) \\
$\bar{x}_\mathrm{mod}$ & \phantom{00}9.8 & \phantom{0}10.5 & \phantom{0}10.9 & \phantom{0}11.3 & \phantom{0}12.1 & \phantom{0}11.3 (0.73)\\
$\bar{x}_\mathrm{heavy}$ & \phantom{0}11.6 & \phantom{0}12.7 & \phantom{0}13.4 & \phantom{0}13.9 & \phantom{0}15.1& \phantom{0}13.7 (0.65) \\
\hline
\end{tabular*}
\end{table}

\subsection{First drinking day}\label{sec7.2}
Another common quantity of interest in alcohol research is the time
until a subject's first drink, also known as the first drinking day, or
FDD. This variable is often treated as the primary outcome in clinical
trials for AUD treatments. We performed a posterior predictive check of
the mean and variance of this statistic across subjects using the same
steps as are outlined in Section~\ref{sec7.1}. This check
relates not only to the heterogeneity of behavior across subjects, but
also to the patterns of drinking behavior across time.

\begin{figure}

\includegraphics{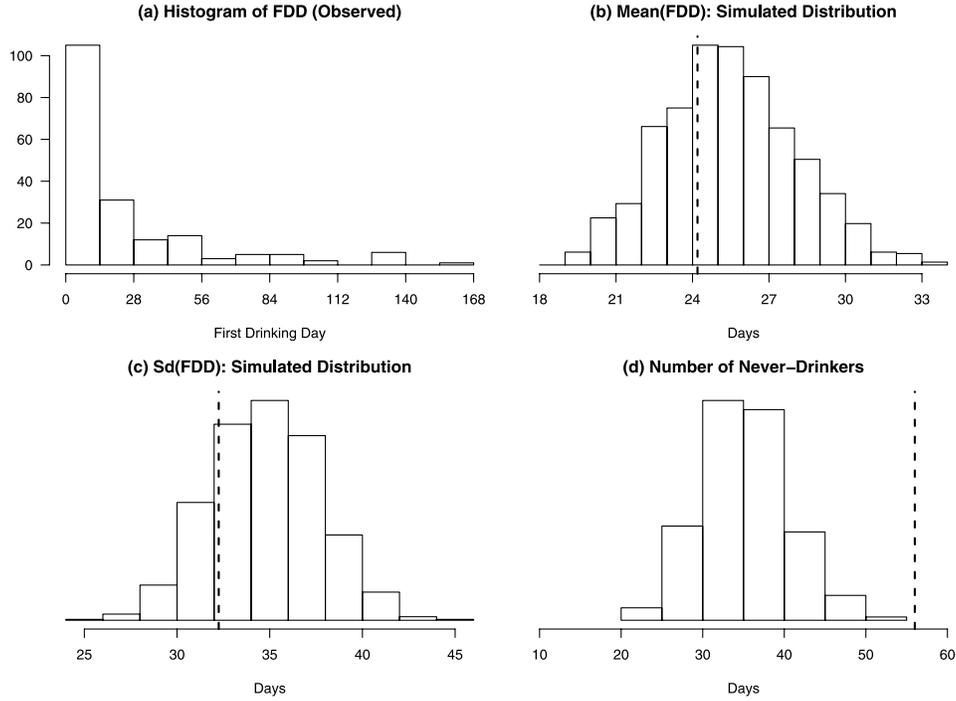}

\caption{\textup{(a)} The~histogram of observed FDD; \textup{(b)} the~histogram of
simulated values of the mean FDD across subjects; \textup{(c)} the~histogram of
simulated values of the standard deviation of FDD across subjects; \textup{(d)}
the~histogram of the simulated distribution of Never-Drinkers. In each
case, the values calculated from the simulated distributions were
computed by ignoring the days that were missing for each subject, so
that the simulated values are directly comparable to the observed
values, denoted by the vertical dotted lines in each figure.}
\label{fig_post_fdd}
\end{figure}

Figure~\ref{fig_post_fdd} contains plots summarizing the posterior
predictive checks of FDD. Figure~\ref{fig_post_fdd}(a) contains a
histogram of the observed values of FDD for all subjects; the empirical
distribution is skewed to the right, and contains only 184 points,
because 56 subjects exhibit a combination of missing data or abstinence
throughout the whole trial. Figures~\ref{fig_post_fdd}(b) and~\ref
{fig_post_fdd}(c) contain histograms of the simulated distributions of
the mean and standard deviation of FDD across subjects, for the subset
of subjects who drank at least once, with dotted lines representing the
observed values. The~observed values lie within the center parts of
their simulated distributions, indicating an adequate fit. Last,
Figure~\ref{fig_post_fdd}(d) contains a histogram of the simulated
values of the number of ``Never-Drinkers'' among the 240 subjects, that
is, those who never drink throughout the course of the clinical trial,
via a combination of abstinence and missing data. The~observed number
of Never-Drinkers is 56 (out of 240), but in general, the model
reproduces data sets in which there are about 20--50 Never-Drinkers.
This indicates a slight lack of fit---there appears to be a set of
individuals among the subjects (who never drink) whose behavior is not
well captured by the model. A similar lack of fit in other settings has
been remedied by the use of a mover-stayer model [Albert (\citeyear{Al1999}), Cook, Kalbfleisch and Yi (\citeyear{CoKaYi2002})], in which a latent class of subjects is constrained to
exhibit constant behavior over the whole time series, and membership in
this class is estimated from the data. On the other hand, it may be
clinically justifiable to suggest that a subject always has a nonzero
probability of drinking, even if in a given sample of days he or she
never drinks. In the fit of the HMM(3), the prior distributions on the
random effects means ($\bolds{\mu}$) virtually prevent any subject from
having a transition probability of zero by design. Transition
probabilities can be very small, but never zero.

\subsection{Patterns across time}\label{sec7.3}
Further posterior predictive checks demonstrate that the HMM(3) is able
to model nonlinear, nonstationary and heteroscedastic drinking
behavior. To see this, we divided the trial period of 24 weeks into 6
blocks of 4-week time periods, and performed posterior predictive
checks on the mean and variance of abstinent, moderate and heavy
drinking days within each time period. In other words, we recreated the
four posterior predictive checks illustrated in Table~\ref{tab_pp} for
data in each of six time periods, and extended the analysis to include
the mean and standard deviation of abstinent days across subjects for
each time period.

\begin{figure}

\includegraphics{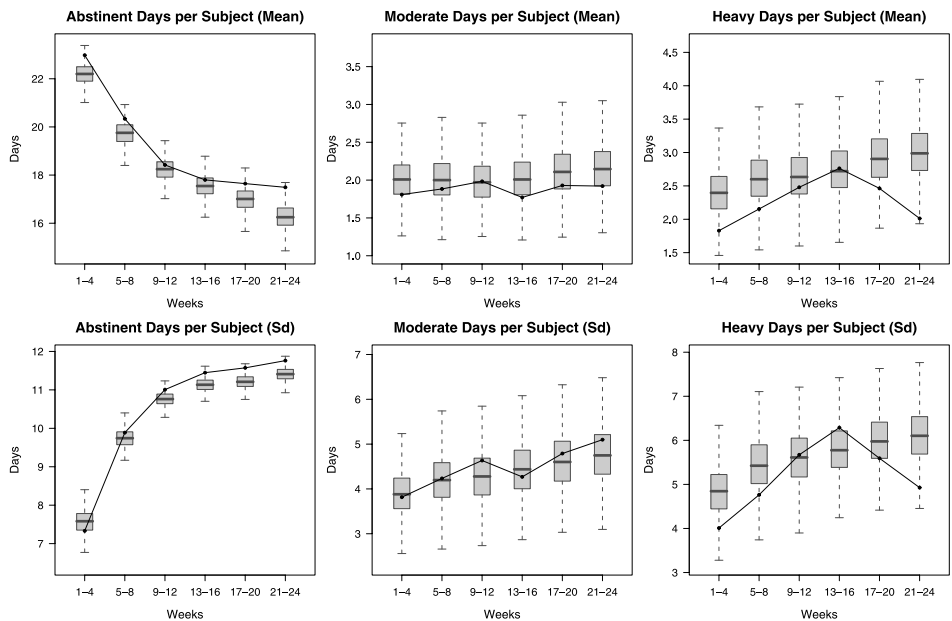}

\caption{Posterior predictive checks of the mean and standard
deviation of all three drinking outcomes in each of the six 4-week time
periods during the trial. The~gray shaded boxplots display the
distributions of simulated values of each statistic, and the black
lines display the observed values of these statistics. For example, for
the first four weeks of the trial, the mean number of abstinent days
across subjects was about 23 days, and in 1000 post-burn-in
simulations, the value of this statistic ranged from approximately
21--24 days, indicating no major lack of fit with respect to this statistic.}
\label{fig_boxplots}
\end{figure}

Figure~\ref{fig_boxplots} contains a visual summary of all 36 of these
posterior predictive checks, where the $x$-axis in each of the six
plots is time. The~boxplots in each of the six plots display the
distributions of the simulated values of the given statistics during
each time period, and the black lines display the observed values of
these statistics. The~plot in the upper left corner, for example, shows
that the observed mean of abstinent days (across subjects) decreased
over the course of the clinical trial, from about 23 days out of a
possible 28 days in the first 4-week time period to about 18 out of 28
days in the last 4-week time period, and that this decreasing trend is
nonlinear. The~HMM(3) captures this nonlinear trend, with a slight
underestimate of the mean number of abstinent days toward the end of
the trial (in the last 4--8 weeks). Furthermore, the lower left-hand
plot shows that the standard deviation of the number of days of
abstinence (across subjects) increases as the trial progresses, also in
a nonlinear way. The~HMM(3) adequately models this heteroscedasticity,
with only a slight underestimate of the standard deviation of abstinent
days in the last 8--12 weeks. Regarding moderate and heavy drinking, the
model appears to adequately describe both the mean and standard
deviation of moderate drinking days for each time period, but fails to
fully capture the apparent quadratic trend in the mean and standard
deviation of heavy drinking through time (both of these quantities
appear to increase from the beginning of the trial until the middle of
the trial, and then decrease until the end of the trial). The~observed
values of these statistics, though, never fall outside of the tails of
the simulated distributions, indicating no major lack of fit. These
posterior predictive checks illustrate the flexibility of the HMM as it
is formulated in Section~\ref{sec3.1}. Note that for categorical
outcomes, the mean and sd are determined by the same sets of
parameters, and that the observed percentages of each categorical
outcome are negatively correlated by definition (because more
realizations of one outcome results in fewer realizations of each of
the other outcomes), so that the posterior predictive checks in
Figure~\ref{fig_boxplots} are not independent.

\subsection{Serial dependence comparison}\label{sec7.4}

To analyze the difference between the Markov model and the HMM(3), we
look for patterns among the data in which one model fits better than
the other. One striking such pattern is related to the way that the two
models handle the serial dependence in the data. Their difference is
apparent in the posterior distributions of the estimated probabilities
(or, equivalently, the deviances) of certain data points. The~HMM(3)
fits better than the Markov model to the third data point in sequences
of the form $(3,1,3)$, $(1,2,1)$, $(2,3,2)$, etc., where two days of equal
consumption are interrupted by a single day of a different level of
consumption. The~Markov model, on the other hand, fits better than the
HMM(3) to the third data point in sequences of the form $(3,1,1)$,
$(1,2,2)$ and $(2,3,3)$, etc., where the second and third day's consumption
levels are equal to each other and different from the first day's consumption.

\begin{figure}[b]

\includegraphics{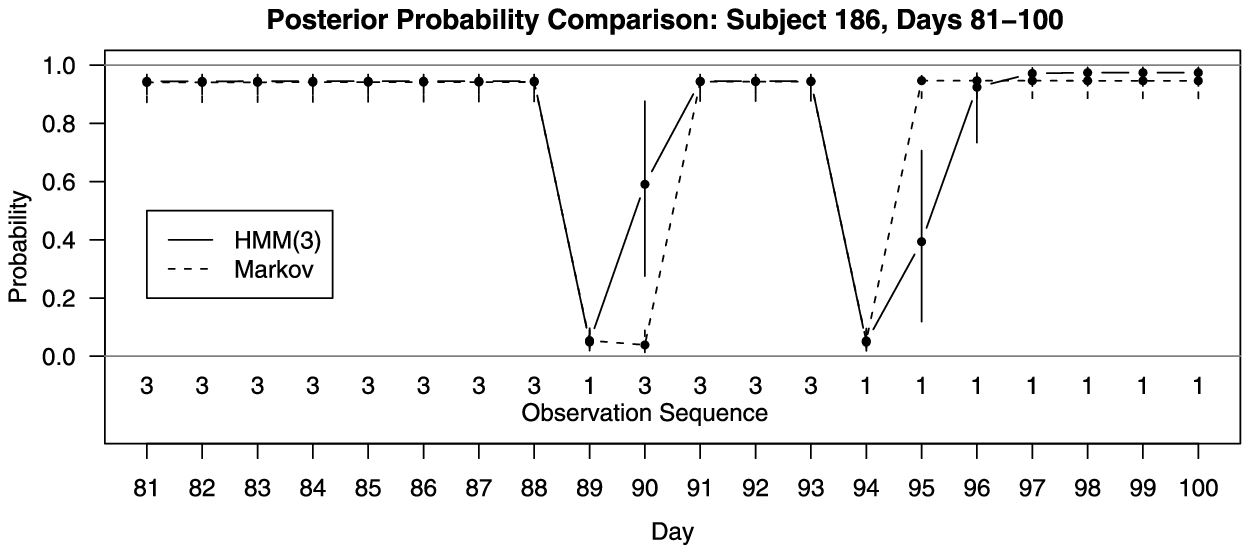}

\caption{Point estimates and vertical 95\% posterior intervals for the
estimated probabilities of a sequence of data points for Subject 186
under both models, where the Markov model point estimates and intervals
are drawn with dashed lines. Both models essentially fit equally well
to the data in this sequence except for two points: the HMM(3) fits
better to the heavy consumption on day 90, and the Markov model fits
better to the abstinence on day 95.}
\label{fig_post_prob}
\end{figure}

Consider Figure~\ref{fig_post_prob}, in which 95\% intervals and point
estimates of the posterior probabilities of a sequence of outcomes for
Subject 186 are plotted for both the HMM(3) and the Markov model.
On day 90, the third day of a $(3,1,3)$ sequence, the HMM(3) fits the
data better, because the subject's reported abstinence on day~89 did
not necessarily represent a change in the hidden state during this
sequence. On day 95, on the other hand, the Markov model fits the data
better, because the reported abstinence that day was the first of many
days in a prolonged sequence of abstinent behavior. In this case both
models fit poorly to the unexpected day of abstinence on day 94, but
the Markov model predicts the following day of abstinence with high
probability, whereas the HMM(3) does more smoothing across time, and is
therefore ``slower'' to predict the second abstinent day (on day~95).
This pattern is generally true for all triplets of the form $(i,j,i)$
and $(i,j,j)$, where $i$ and $j$ represent observed consumption levels.

In the presence of measurement error---which is plausible given that
the subjects self-report their alcohol consumption---and in light of
the cognitive behavioral model of relapse, which suggests that a change
in one's alcohol consumption is not necessarily equivalent to a change
in one's underlying health state, the ``smoother'' fit of the HMM(3) is
an advantage over the Markov model. The~notion that the underlying
behavior of individuals with an AUD is relatively steady, and that the
observations are somewhat noisy, is consistent with the cognitive
behavioral model of relapse.

\subsection{Differences across subjects}\label{sec7.5}
To see the different drinking behaviors across subjects, we simulated
sample paths from the fitted model for each subject using a procedure
similar to the 4-step procedure outlined in Section~\ref
{sec7.1}. This time, however, instead of simulating new
random effects, $\bolds{\alpha}^\mathrm{rep}$, from the posterior
distributions of $\bolds{\mu}$ and $\bolds{\sigma}$ to represent a
new sample of
subjects from the population, we simply used the posterior samples of
$\bolds{\alpha}$ directly to simulate new hidden states, $\mathbf
{H}^\mathrm{rep}$,
and observations, $\mathbf{Y}^\mathrm{rep}$ (steps 2 and 3 of the procedure).
This way each set of simulated time series represents a new set of
realizations for the exact same set of subjects whose data were
observed, and we can make direct comparisons between the observed data
and the simulations for each individual.

\begin{figure}

\includegraphics{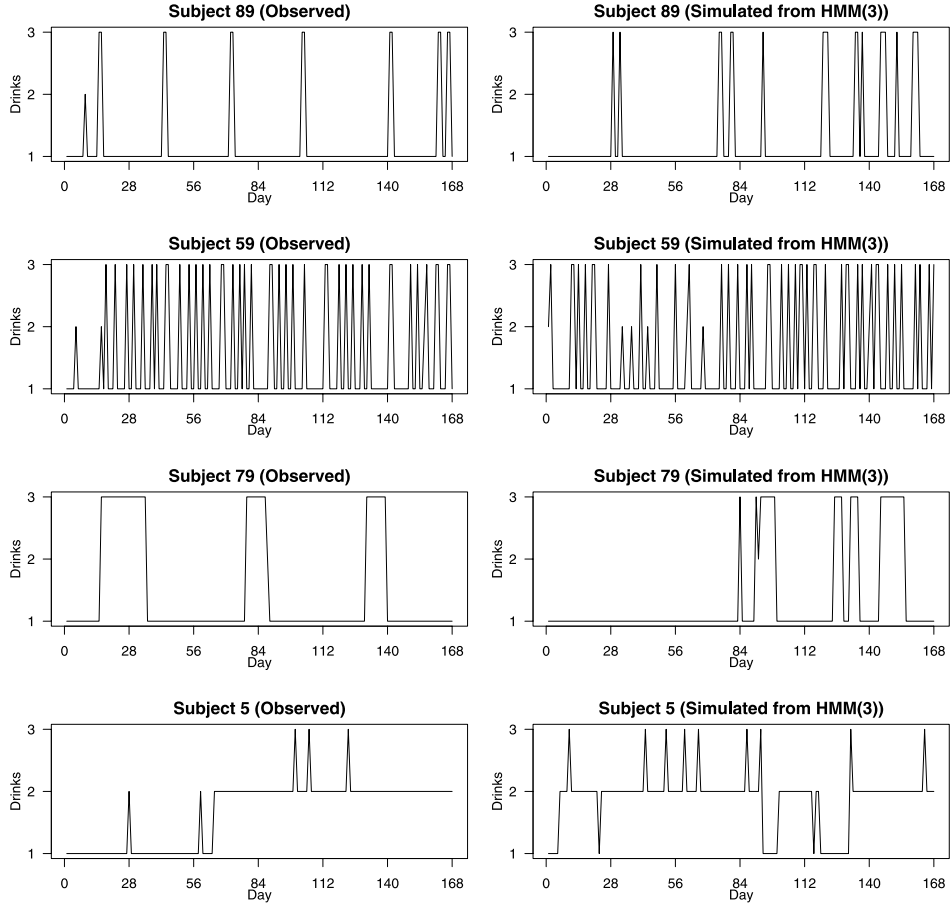}

\caption{A visual comparison of observed (left column) and simulated
(right column) time series for four subjects whose behavior varied
widely. The~HMM(3) is able to reproduce a wide variety of behaviors via
its random effects in the hidden state transition matrix.}\vspace*{5pt}
\label{fig_paths}
\end{figure}

Figure~\ref{fig_paths} displays the observed drinking time series and
one randomly drawn simulated time series for four subjects whose
behaviors varied widely from each other. From the figure it is clear
that the random effects in the HMM(3) allow it to capture a wide
variety of drinking behaviors. Here is a summary of these four drinking
behaviors:
\begin{enumerate}
\item Subject 89 is almost always abstinent, with occasional days of
heavy drinking,
\item Subject 59 drinks heavily in very frequent short bursts
(typically for just \mbox{1--2}~days in a row),
\item Subject 79 drinks heavily on occasion for longer periods of time
(10 days in a row appears typical),
\item Subject 5 drinks moderately for prolonged periods of time.
\end{enumerate}
Each of these four unique drinking behaviors is captured by the HMM(3)
through the random effects in the hidden state transition matrix.

\section{A new definition of relapse}\label{sec8}

The~HMM offers a new definition of a relapse: any time point at which
the probability is high that a subject is in hidden state 3, the Heavy
Drinking state. There are a variety of ways to estimate the hidden
states in an HMM, many of which are discussed in Scott (\citeyear{Sc2002}). Most
methods can be categorized in one of two ways: They either maximize the
marginal probability of each $H_{it}$ separately, or they attempt to
find the set of sequences of hidden states $\mathbf{H}$ (which we
think of as
an $N\times T$ array, consisting of a set of $N$ sequences, each of
length $T$) that is collectively the most likely (i.e., the mode of the
joint posterior distribution of $\mathbf{H}$). The~first method
requires the
calculation of the marginal probabilities of each hidden state at each
time point, and then estimates the hidden state for each subject and
time point as $\max_s\{\mathrm{P}(H_{it}=s \vert \mathbf{Y}, \bolds
{\theta})\}$.
When the hidden state transition matrix contains zeroes, though, or has
high probabilities on the diagonal (as it does in our case for many
individuals), then using the marginal probabilities to estimate hidden
states can result in a sequence that is very unlikely, or even
impossible, because maximizing the marginal probabilities does not
fully account for the dependence between consecutive hidden states
[Rabiner and Juang (\citeyear{RaJu1986})]. Such a sequence is not appealing for
characterizing relapse periods because our goal is to segment a
subject's drinking time series into plausible periods of in-control
drinking behavior as opposed to out-of-control (relapse) drinking behavior.

We attempt the second type of hidden state estimation, in which we seek
to find the set of sequences $\mathbf{H}$ that is the most likely,
given the
observed data. In the Bayesian framework, one way to do this would be
to select the sequence $\mathbf{H}^{(g)}$ that occurs most often in our
post-burn-in posterior samples $g=1,\ldots ,G$. With such a large data set,
though, it would require a huge number of MCMC iterations to ensure
that a single realization of $\mathbf{H}$ was sampled twice or more.
In our
posterior sample, for example, the largest amount of overlap between
any two samples of $\mathbf{H}$ was about 93\% [that is, the pair of sets
$(\mathbf{H}^{(i)},\mathbf{H}^{(j)})$ for $i,j \in(1,\ldots ,G)$ that
had the most
elements in common shared about 93\% of their elements]. Thus, to
estimate $\mathbf{H}$, we used a different method: we computed the most
likely set of sequences given the observed data and the posterior mean,
$\bar{\bolds{\theta}}$, using the Viterbi algorithm [Rabiner and Juang
(\citeyear{RaJu1986})]. This is similar to what is done is most frequentist analyses
of HMMs, in which the Viterbi algorithm is run conditional on the MLEs
of the model parameters to compute the most likely sequence(s) of
hidden states.

\begin{figure}

\includegraphics{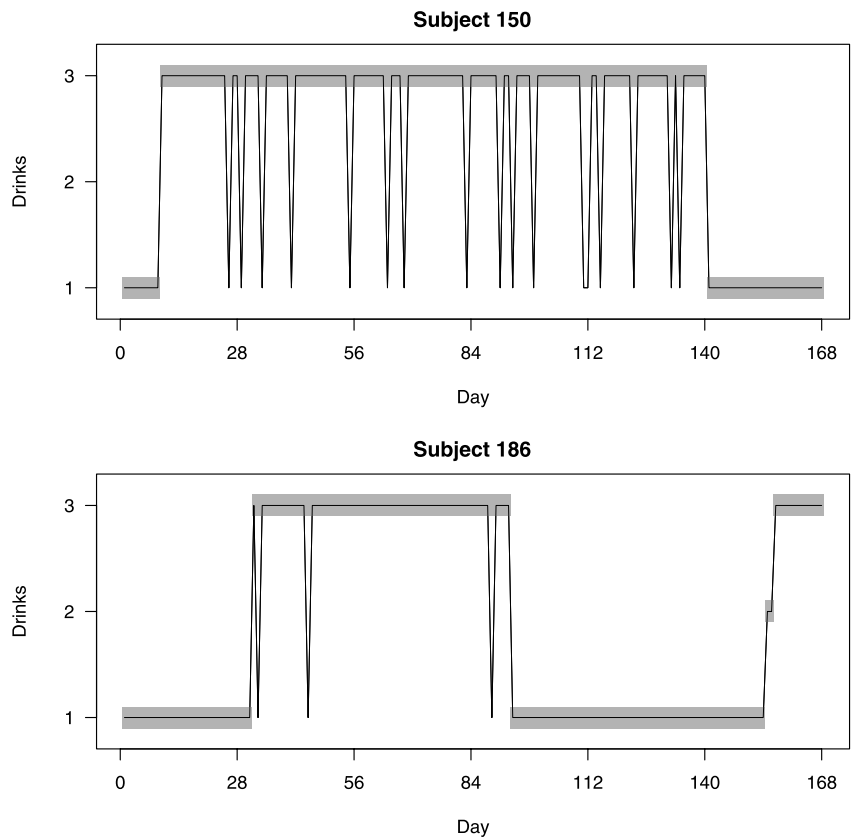}

\caption{The~most likely sequence of hidden states for Subjects 150
and 186. The~dark lines represent the observed daily drink counts, and
the shaded regions represent the estimated hidden state at each time
point, where the hidden states are numbered 1 through 3 as they are in
Table~\protect\ref{tab_cond_dists}.}
\label{fig_viterbi}
\end{figure}

Figure~\ref{fig_viterbi} shows the observed data and the most likely
hidden states for two subjects. Subject 150 is estimated to be in the
Heavy Drinking state from about Day 10 to about Day 140, despite not
drinking on a number of days during that span. Likewise, Subject 186 is
estimated to be in the Heavy Drinking state for a prolonged period of
time despite a few days of abstinence during that time span. According
to the definition of relapse that says any day spent in the Heavy
Drinking hidden state is a relapse, these subjects were both in a state
of relapse for a long, continuous period of time. This relapse
definition is robust to outliers and measurement errors in the sense
that a single day, or a brief ``burst,'' among the observations does not
necessarily imply a transition in the most likely hidden state
sequence. On many days, these two subjects reported a single day of
abstinence, preceded and followed by many consecutive days of heavy
drinking. Some definitions of relapse would be sensitive to these
abstinence observations, and would suggest that the subject was no
longer in a state of relapse, whereas the HMM-based definition of
relapse includes these days as well, because the most likely hidden
state sequence remains the same whether those observations were a 1, 2
or~3, because of the dependence structure across time that the HMM imposes.\looseness=1

Using this definition of relapse, a relapse depends on one's
unobservable physical and mental state of health, rather than directly
on one's drinking behavior. Choosing which relapse definition is best
for a given purpose requires clinical insight. Currently used
definitions of relapse define what patterns of drinking constitute a
relapse a priori before looking at the data. There is no consensus on
what patterns constitute a relapse [Maisto et al. (\citeyear{MaPoCoLyMa2003})]. An advantage
of the HMM approach is that it starts with a model of drinking behavior
and a qualitative definition of relapse (e.g., a state in which a
subject has a substantial probability of drinking heavily) and then
lets the data decide what patterns of drinking constitute a relapse.
This being said, we found evidence from earlier fits of HMMs to this
data that the HMM-based definitions of relapse are sensitive to the
number of hidden states in the model and the choice of which hidden
states are categorized as relapse states. This choice becomes harder as
the number of hidden states grows, because with more hidden states, the
conditional distributions tend to correspond less clearly to easily
identifiable drinking behaviors. Thus, using the HMM definition of
relapse suggested in this section still requires careful thought on the
part of the investigator with respect to (1), the choice of the number
of hidden states, and, once that choice is made, to (2), the choice of
which hidden states will be designated as relapse states.

\section{Discussion and extensions}\label{sec9}

In this paper we develop a nonhomogenous, hierarchical Bayes HMM with
random effects for data from a clinical trial of an alcoholism
treatment. The~model is motivated by data which exhibits flat stretches
and bursts, as well as by the cognitive-behavioral model for relapse
that coincides with the structure of the HMM, in which subjects make
transitions through unobservable mental/physical states over time, and
their daily alcohol consumption, which is observed, depends on their
underlying state. A major strength of the HMM (and also of the Markov
model we propose) is that it models the daily alcohol consumption of
subjects. Such a model provides a rich description of subjects' alcohol
consumption behavior over time, as opposed to some lower-dimensional
random variable such as the time until a subject's first relapse, which
provides a narrower view of a subject's alcohol consumption.

The~fit of the model to the data reveals a number of interesting
things. The~HMM with three hidden states fits well, and the hidden
state transition matrix reveals that the most persistent state is
abstinence, followed by heavy drinking, followed by moderate drinking.
The~transition matrices, however, vary widely among subjects, and the
models, in general, are able to reproduce a wide variety of drinking
behaviors (see Figure~\ref{fig_paths}). The~treatment in the clinical
trial we analyzed, Naltrexone, appears to have a small, beneficial
effect on certain transition probabilities, as well as on the overall
percentage of time one would expect to spend in the heavy drinking
state. Posterior predictive checks reveal that the heterogeneity among
individuals is captured by the random effects in the hidden state
transition matrices. Furthermore, the HMM(3) adequately models
nonstationary, nonlinear and heteroscedastic drinking patterns across time.

Also, the HMM suggests a new definition of relapse, based on hidden
states, which is conceptually supported by the cognitive-behavioral
model of relapse, and can be compared to existing definitions of
relapse to offer new insights into which patterns of drinking behavior
are indicative of out-of-control behavior (i.e., a substantial
probability of heavy drinking).

\begin{appendix}

\section*{Appendix: MCMC algorithms to fit the models}\label{section_appendix}
Here we present some details of the MCMC algorithms used to fit the models.

\subsection{Fitting the Markov model}
The~Markov model, as was described in Section \ref{sec3.2}, is
essentially composed of $M$ separate multinomial logit models, one for
each row of the transition matrix. The~Markov model, however, is
substantially more difficult to fit, because the observations given the
model parameters are not i.i.d.; each observation depends on the
previous one. We incorporate a data augmentation step into the
Metropolis-within-Gibbs algorithm to fit the Markov model. First, the priors:
%
\begin{eqnarray}
\quad \beta_{mjk} &\sim& N(0,10^2)\qquad  \mbox{for } m=1,\ldots ,M,
j=2,\ldots ,M, k=1,\ldots ,p, \label{eqn_markov_prior_beta}\\
\quad \mu_{mj} &\sim& N(0,10^2)\qquad  \mbox{for } m=1,\ldots ,M,
j=2,\ldots ,M, \label{eqn_markov_prior_mu} \\
\quad p(\sigma_{mj}) &\sim& 1 \qquad  \mbox{for } m=1,\ldots ,M,
j=2,\ldots ,M ,\label{eqn_markov_prior_sigma} \\
\quad \bolds{\pi}&\sim& \operatorname{Dirichlet}(\mathbf{1}). \label
{eqn_markov_prior_pi}
\end{eqnarray}

We fit the Markov model with the following steps:
\begin{enumerate}
\item Initialize the parameters, $\alpha_{ijm}^{(0)}, \bolds{\beta}
_{jm}^{(0)} $, for $i=1,\ldots ,N$, $j=1,\ldots ,M$, and\vspace*{-2pt} $m=2,\ldots ,M$, using
overdispersed starting values, and initialize $\mu_{jm}^{(0)}=0$ and\vspace*{-2pt}
$\sigma_{jm}^{(0)}=1$ for $j=1,\ldots ,M$, and $m=2,\ldots ,M$.\vspace*{1pt}
\item For $g=1,\ldots ,G$,
\begin{enumerate}[(c)]
\item[(a)] Sample $\mathbf{Y}_i^{\mathrm{mis},(g)} \vert \mathbf{Y}_i^{\mathrm{obs}},
\bolds{\alpha}_i^{(g-1)}, \bolds{\beta}^{(g-1)}$ from its full conditional
distribution for $i=1,\ldots ,N$.
\item[(b)] Sample $\alpha_{ijm}^{(g)} \vert \mathbf{Y}_i^{\mathrm{obs}},
\mathbf{Y}
_i^{\mathrm{mis},(g)}, \bolds{\beta}_{j}^{(g-1)}, \mu_{jm}, \sigma_{jm}$
using a univariate random walk Metropolis sampler for $i=1,\ldots ,N$,
$j=1,\ldots ,M$, and $m=2,\ldots ,M$.
\item[(c)] Sample $\beta_{jmk}^{(g)} \vert \mathbf{Y}^{\mathrm{obs}},
\mathbf{Y}^{\mathrm
{mis},(g)}, \bolds{\alpha}_{j}^{(g)}$ using a univariate random-walk
\break
Metropolis sampler, for $j=1,\ldots ,M$, $m=2,\ldots ,M$, and $k=1,\ldots ,p$, where
$\bolds{\alpha}_{j}^{(g)}$ denotes the $N\times (M-1)$ matrix of random
intercepts sampled during iteration $g$.
\item[(d)] Sample $\mu_{jm}^{(g)} \vert \bolds{\alpha}_{jm}^{(g)}, \sigma
_{jm}^{(g-1)}$ from its full conditional normal distribution for\vspace*{-2pt}
$j=1,\ldots ,M$ and $m=2,\ldots ,M$, where $\bolds{\alpha}_{jm}^{(g)}$ is the
length-$N$ vector of random intercepts sampled during iteration $g$.
\item[(e)] Sample $\sigma_{jm}^{(g)} \vert \bolds{\alpha}_{jm}^{(g)}, \mu
_{jm}^{(g)}$ from its full conditional inverse chi-squared distribution
for $j=1,\ldots ,M$ and $m=2,\ldots ,M$.
\item[(f)] Sample $\bolds{\pi}\vert \mathbf{Y}^{\mathrm{obs}}, \mathbf
{Y}^{\mathrm{mis},(g)}$
from its full conditional Dirichlet distribution.
\end{enumerate}
\end{enumerate}

Given the full set of observations for each subject $(\mathbf
{Y}_i^{\mathrm
{obs}}, \mathbf{Y}_i^{\mathrm{mis}})$, which we have after step (a) of each
iteration, we easily sample the model parameters [steps~\mbox{(b)--(e)} in the
algorithm] associated with each row $j$ of the transition matrix by
gathering all the observations $\{Y_{it} \in\mathbf{Y}_{(i,1:(T-1))} \dvtx
Y_{it}=j\}$, and then treating the subsequent observations,
$Y_{(i,t+1)} \in(1,2,\ldots ,M),$ as draws from a multinomial distribution
with probability distributions given in Equation (\ref{model_markov}),
and we do this for rows $j=1,\ldots ,M$.

\subsection{Fitting the HMM}
The~priors for the HMM are the same as they are for the Markov model,
except we replace $M$ with $S$ in Equations~(\ref
{eqn_markov_prior_beta})--(\ref{eqn_markov_prior_pi}), and we add the
prior distributions for the parameters in the conditional
distributions, $\mathbf{P}_s \sim\operatorname{Dirichlet}(\mathbf{1})$, for
$s=1,\ldots ,S$.

The~HMM fit requires a similar, but slightly different
Metropolis-within-Gibbs MCMC algorithm from that of the Markov model.
We follow Scott (\citeyear{Sc2002}) and, in each iteration of the MCMC algorithm,
use a forward-backward recursion to evaluate the likelihood, sample the
hidden states given the parameters, and then sample the parameters
given the hidden states. The~details are as follows:
\begin{enumerate}
\item First, initialize all model parameters. For the HMM, we
initialize the parameters in a specific way so as to avoid the
label-switching problem that is common to fitting HMMs, and can be
triggered by widely dispersed starting points. First, we fit the HMM
with $S$ hidden states using the EM algorithm, without incorporating
covariates and with complete pooling across subjects. We then set the
values of $\bolds{\beta}_{rs}^{(0)}$ to zero, and set the values of
$\mu
_{rs}^{(0)}$ to the values such that, if all the $N$ random intercepts
$\alpha_{irs}^{(0)}$, for $i=1,\ldots ,N$, were set to equal $\mu
_{rs}^{(0)}$, the model would have the same exact transition
probabilities as MLE transition probabilities estimated from the EM
algorithm. This usually ensures that each chain in a multichain MCMC
algorithm converges to the same mode of the posterior distribution,
which has $S!$ symmetric modes corresponding to the permutations of the
state labels.
\item For $g=1,\ldots ,G$,
\begin{enumerate}[(c)]
\item[(a)] Sample the vector $\mathbf{H}_i \vert \mathbf{Y}_i^\mathrm{obs},
\bolds{\beta},
\bolds{\alpha}_i, \mathbf{P}, \bolds{\pi}$ from it's full
conditional distribution using
the forward-backward recursion as described in detail in Scott (\citeyear{Sc2002}),
for $i=1,\ldots ,N$.
\item[(b)] Sample $\bolds{\beta}$, $\bolds{\alpha}$, $\bolds{\mu}$ and
$\bolds{\sigma}$ given $\mathbf{H}$
just as they are sampled in steps (b)--(e) in the Markov model, except
replace the indices $j$ and $m$ with $r$ and $s$, respectively, for
$r=1,\ldots ,S$ and $s=2,\ldots ,S$, and replace $(\mathbf{Y}^{\mathrm
{obs}},\mathbf{Y}
^{\mathrm{mis}})$ with $\mathbf{H}$.
\item[(c)] Sample $\mathbf{P}\vert \mathbf{H}, \mathbf{Y}^\mathrm{obs}$
from its full
conditional distribution.
\item[(d)] Sample $\bolds{\pi}\vert \mathbf{H}_1$ from its full
conditional Dirichlet
distribution, where $\mathbf{H}_1$ denotes the length-$N$ vector of hidden
states at time $t=1$.
\end{enumerate}
\end{enumerate}

Data Files
Included in the supplementary materials are two data
files. The~first file, ``y.csv,'' contains the ordinal drink counts for
each subject on each day, and has $N=240$ rows and $T=168$ columns. The~second file, ``x.csv,'' contains the covariates used to fit our models,
and contains sex, treatment and prior drinking behavior for each subject.
\end{appendix}

\begin{supplement}
\stitle{Data Files}
\slink[doi]{10.1214/09-AOAS282SUPP}
\slink[url]{http://lib.stat.cmu.edu/aoas/282/supplement.zip}
\sdatatype{.zip}
\sdescription{Included in the supplementary materials are two data
files. The~first file, ``y.csv,'' contains the ordinal drink counts for
each subject on each day, and has $N=240$ rows and $T=168$ columns. The second file, ``x.csv,'' contains the covariates used to fit our models,
and contains sex, treatment and prior drinking behavior for each subject.}
\end{supplement}

\section*{Acknowledgments}
The~authors thank Andrew Gelman and Matthew Scho\-field for their helpful
comments on earlier drafts of this paper, as well as the Associate
Editor and three anonymous referees whose comments greatly improved the paper.

\printaddresses

\end{document}